\documentclass[final,5p,times,twocolumn]{elsarticle}
\usepackage{graphicx}
\usepackage{epsfig}

\usepackage{amssymb}
\usepackage{amsthm}

\journal{Physics Letters B}

\begin{document}

\begin{frontmatter}



\title{Nuclear coherent population transfer with x-ray laser pulses}


\author{Wen-Te Liao}
\ead{Wen-Te.Liao@mpi-hd.mpg.de}

\author{Adriana P\'alffy \corref{cor1}}
\ead{Palffy@mpi-hd.mpg.de}

\author{Christoph H. Keitel}
\ead{Keitel@mpi-hd.mpg.de}

\address{Max-Planck-Institut f\"ur Kernphysik, Saupfercheckweg 1, D-69117 Heidelberg, Germany}
\cortext[cor1]{Corresponding author. Tel.: +49(0)6221 516162, fax: +49(0)6221 516152}

\begin{abstract}

Coherent population transfer between nuclear states using x-ray laser pulses is studied. The laser pulses drive two nuclear 
transitions between three nuclear states in a setup reminding of stimulated Raman adiabatic passage used for atomic coherent population transfer.  To compensate for the lack of $\gamma$-ray laser sources, we envisage accelerated nuclei interacting with two copropagating or crossed x-ray laser pulses. The parameter regime for nuclear coherent population transfer using fully coherent light generated by future X-Ray Free-Electron Laser facilities and moderate or strong acceleration of nuclei is determined. We find that the most promising case requires laser intensities of $10^{17}$-$10^{19}$ W/cm$^{2}$ for complete nuclear population transfer. As  relevant application, the controlled pumping or release of energy stored in long-lived nuclear states is discussed.

\end{abstract}


\end{frontmatter}


Long-lived excited nuclear states, also known as isomers, can store large amounts of energy over longer periods of time. Isomer depletion, i.e., release on demand of the energy stored in the metastable state, has received  great attention in the last one and a half decades, especially related to the fascinating prospects of nuclear batteries \cite{PhilWalker1999,K.W.D.Ledingham2003,Sun2005,isomersPRL}. Depletion occurs when the isomer is excited to a higher level, which is
associated with freely radiating states and therefore releases
the energy of the metastable state. Coherent population transfer between nuclear states would therefore not only be a powerful tool for preparation and detection in nuclear physics, but also especially useful for control of energy stored in  isomers.

In atomic physics, a successful and robust way for atomic coherent population transfer is 
the stimulated Raman adiabatic passage  (STIRAP) \cite{K.Bergmann1998}, a technique in which two coherent fields couple to a three-level system. The transfer of such schemes to nuclear systems, although encouraged by progress of laser technology, has not been accomplished due to the lack of $\gamma$-ray laser sources. The pursuit of coherent sources for wavelengths around or below 1~\AA \ is
supported however by the advent and commissioning of x-ray
free electron lasers, the availability of which will stimulate the transfer of quantum optical schemes to nuclei.

To bridge the gap between x-ray laser frequency and  nuclear transition energies, a  key proposal is to combine  moderately accelerated target nuclei and novel x-ray lasers \cite{ThomasJ.Buervenich2006}. Using this scenario, the interaction of x-ray from the European X-ray Free Electron Laser (XFEL) \cite{xfel} with  nuclear two-level systems was studied theoretically \cite{ThomasJ.Buervenich2006,P'alffy2008}.
The manipulation of nuclear state population by STIRAP and the coherent control of isomers have however never been addressed, partially because of the poor coherence properties of the XFEL.

\begin{figure}[h]
\center{
  \includegraphics[width=8cm]{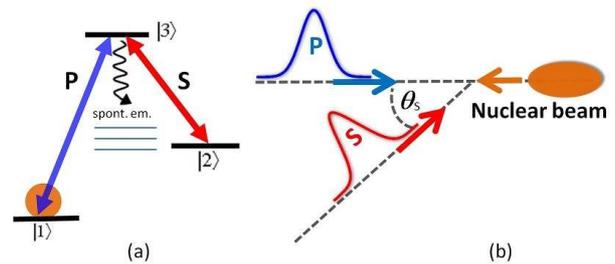}}
  \caption{\label{fig1}
  (a) The nuclear $\Lambda$-scheme. The initial population is concentrated in state $\vert 1\rangle$. The pump laser $P$ drives the transition $\vert 1\rangle\rightarrow\vert 3\rangle$, while the Stokes laser $S$ drives the transition $\vert 2\rangle\rightarrow\vert 3\rangle$. The upper state $\vert 3\rangle$  decays also to other states through spontaneous emission. (b) STIRAP: two partially overlapping  x-ray laser pulses $P$ (pump) and $S$ (Stokes) interact with relativistically accelerated nuclei. The collinear beams setup corresponds to $\theta_S=0$.
  }
\end{figure}

In this Letter we investigate for the first time the nuclear coherent population transfer (NCPT) between the two lower states in the nuclear $\Lambda$-level scheme showed in Fig.~\ref{fig1}(a) using two overlapping x-ray laser pulses in a STIRAP setup. This is a typical  three-level scheme that can lead to the depletion of a metastable state, here the ground state $|1\rangle$, via a triggering level $|3\rangle$ to a level $|2\rangle$ whose decay to the nuclear ground state is no longer hindered by the long-lived isomer. We show that using a fully coherent XFEL such as the future XFEL Oscillator (XFELO) \cite{Kim2008} or the seeded 
XFEL (SXFEL) \cite{sxfel.Feldhaus,sxfel.Saldin,xfel,slac,LCLSII} for both pump and Stokes lasers, together with acceleration of the target nuclei to achieve the resonance condition, allow for NCPT.  The coherence of the x-ray laser has as a result  nuclear coherent control at much lower intensities than previous calculated values for laser driving of nuclear transitions \cite{ThomasJ.Buervenich2006}, already at  $10^{17}$-$10^{19}$ W/cm$^{2}$. In view of our results, the indeed challenging experimental prospects of isomer depletion are discussed and a setup to produce both pump and Stokes pulses with different frequencies in the nuclear rest frame from a single coherent x-ray beam is put forward. 

The interaction of a nuclear $\Lambda$-level scheme with the pump laser $P$ driving the $\vert 1\rangle\rightarrow\vert 3\rangle$  transition and the Stokes laser $S$ driving the $\vert 2\rangle\rightarrow\vert 3\rangle$  transition is depicted in Fig.~\ref{fig1}(a).  In STIRAP, at first the Stokes laser  creates a superposition of the two  unpopulated states $\vert 2 \rangle$ and $\vert 3 \rangle$.  Subsequently, the pump laser couples the fully occupied 
$\vert 1 \rangle$ and the pre-built coherence of the two empty states. The dark (trapped) state is formed and evolves with the time dependent Rabi frequencies of the pump and Stokes fields $\Omega_{p}$ and $\Omega_{S}$,  respectively \cite{K.Bergmann1998}. 

Typically,  the $\Lambda$-level scheme is not closed, i.e. the population in $\vert 3\rangle$ will not only decay to $\vert 1\rangle$ and $\vert 2\rangle$ but also to other low energy levels through  spontaneous radiative decay or by other decay mechanisms such as internal conversion or $\alpha$ decay. This open feature of $\vert 3\rangle$ speaks against direct  pumping, allowing us to identify two situations: 
$(i)$ the lifetime of $\vert 3\rangle$ is longer than the pulse duration. Since the nucleus can stay in $\vert 3\rangle$ long enough, apart from STIRAP, also NCPT via sequential isolated pulses such as  $\pi$ pulses, i.e. pulses that transfer the complete nuclear state population from one state to another,  is possible. A first $\pi$ pulse can pump the nuclei from $|1\rangle$ to $|3\rangle$, followed by a second Stokes $\pi$ pulse  that drives the $|3\rangle\rightarrow|2\rangle$ decay. The latter scenario lacks the robustness of STIRAP, having a sensitive dependence on the laser intensities. 
$(ii)$ the lifetime of $\vert 3\rangle$ is shorter than the pulse duration. Because of the  high decay rate of $\vert 3\rangle$, separated single pulses cannot produce NCPT and STIRAP  provides the only possibility for population transfer.

The nuclear excitation energies in the two  regimes described above are typically higher than the designed photon energy of the XFELO and SXFEL. Nuclei  suitably accelerated can interact with two Doppler-shifted x-ray laser pulses. The two laser frequencies and the relativistic factor $\gamma$ of the accelerated nuclei have to be chosen such that in the nuclear rest frame both one-photon resonances are fulfilled. Copropagating laser pulses (with $\theta_S=0$ in Fig.~\ref{fig1}(b)) should have different frequencies in the laboratory frame in order to match the nuclear transition energies. To fulfill the resonance conditions with a single-color laser we envisage the pump and Stokes pulses meeting the nuclear beam at different angles ($\theta_S\ne 0$),  as shown in Fig.~\ref{fig1}(b).

In general, situation $(i)$ is related to nuclear excitations of tens up to hundreds of keV, such that $\gamma\lesssim 10$. These low-lying levels have however energy widths of about 1 $\mu$eV or less,  orders of magnitude smaller than the photon energy spread. In this case only a fraction of the incoming photons will drive the nuclear transition, leading to a small effective intensity \cite{P'alffy2008}. For case $(ii)$, the required $\gamma$ for driving  MeV transitions is on the order of $20-100$. Typically, such transitions have widths ($\sim$ 1~eV) larger than the bandwidth of the XFELO or SXFEL. The effective and nominal laser intensity have in this case the same value, an advantage of the high-$\gamma$ regime.  A list of parameters for a number of nuclei with suitable  transitions for both  $(i)$ and $(ii)$  regimes  is presented  in Tables~\ref{table1} and \ref{table2}. 

\begin{table}[h]
\vspace{-0.4cm}
\caption{\label{table1}
Laser and nuclear beam parameters in the laboratory frame. The  accelerated nuclei have the relativistic factor $\gamma$. For the copropagating-beams setup, the Stokes photon energy $E_{S}$ is given in keV.  The pump (copropagating beams) or both pump and Stokes lasers (crossed beams) photon energies are 12.4~keV for SXFEL and 25~keV for XFELO, respectively. For the crossed-beam setup, the angle $\theta_{S}$ between the pump and Stokes depicted in Fig.~\ref{fig1}(b) is given in rad.
}

\center{
\begin{tabular}{r|ccc|ccc}
\hline
 &          \multicolumn{3}{c}{SXFEL} & \multicolumn{3}{c}{XFELO} \\
\cline{2-7}
Nucleus& $\gamma$ & $\theta_{s}$ & $E_{S}$ &   $\gamma$ & $\theta_{s}$ & $E_{S}$\\
\hline
$^{185}$Re &11.5&1.4544&6.93&5.7&1.4596&13.97\\
$^{97}$Tc &22.6&1.3836&7.36&11.2&1.3848&14.83\\
$^{154}$Gd &50.1&0.6407&11.17&24.8&0.6408&22.52\\
$^{168}$Er&72.0&0.4260&11.85&35.7&0.4260&23.88\\
\hline
\end{tabular}
}

\end{table}

We study the dynamics of the system depicted in Fig.~\ref{fig1}(a) in the nuclear rest frame. This 
is governed by the master equation for the density matrix $\widehat{\rho}$ \cite{K.Bergmann1998, Scully2006} that reads $\frac{\partial}{\partial t}\widehat{\rho} = \frac{1}{i\hbar}\left[ \widehat{H},\widehat{\rho}\right]+\widehat{\rho}_{relax}$, with  
the interaction Hamiltonian
\begin{equation}
\widehat{H} = -\frac{\hbar}{2}
\left( 
\begin{array}{ccc}
  0 & 0 & \Omega_{p}^{*}\\
  0 & 2\left(\bigtriangleup_{p}-\bigtriangleup_{S} \right)  & \Omega_{S}^{*}\\ 
  \Omega_{p} & \Omega_{S} & 2\bigtriangleup_{p}
\end{array}  
\right)\, ,
\end{equation}
and the relaxation matrix $\widehat{\rho}_{relax}$ that includes the spontaneous decay.
The initial conditions  are $\rho_{ij}(0) = \delta_{i1}\delta_{1j}$. In the expression above, $\Delta_{p(S)}=\gamma(1+\beta  \cos\theta)\omega_{p(S)}-ck_{31(2)}$ is the laser detuning, where $\gamma$ and $\beta$ denote the relativistic factors, $\gamma=1/\sqrt{1-\beta^2}$, $c$ is the speed of light, $\omega_{p(S)}$ is the pump (Stokes) laser angular frequency and $k_{31}$ and $k_{32}$ are the wave numbers of the corresponding transitions. The angle $\theta$ is zero for the pump laser and $\theta=\theta_S$ for the Stokes laser.
The slowly varying effective Rabi frequencies $\Omega_{p(S)}(t)$ in the nuclear rest frame for nuclear transitions of electric ($\varepsilon$) or  magnetic ($\mu$) multipolarity $L$ are given by \cite{K.Bergmann1998,P'alffy2008}
\[
\Omega_{p(S)}(t)  = \frac{4\sqrt{\pi}}{\hbar}\left[\frac{\gamma^2(1+\beta \cos\theta)^2I^{\mathrm{eff}}_{p(S)}(L+1)B(\varepsilon/\mu\, L)}{c\epsilon_{0}L}\right]^{1/2} 
\]
 \begin{equation}  
   \label{eq4}  
 \times \frac{k_{31(2)}^{L-1}}{(2L+1)!!} \mathrm{Exp}\left\lbrace -\left[\frac{\gamma(1+\beta \cos\theta)(t-\tau_{p(S)})}{\sqrt{2}T_{p(S)}}\right]^{2}\right\rbrace\, . 
\end{equation}
Here we have expressed the nuclear multipole moment  with the help of the reduced transition probabilities $B(\varepsilon/\mu\, L)$ following the approach developed in \cite{P'alffy2008}. This allows for a unified treatment of the laser-nucleus interaction for both dipole-allowed ($E1$) and dipole-forbidden nuclear transitions.  
All the laser quantities have been transformed in Eq.~(\ref{eq4}) into the nuclear rest frame, leading to the angular frequency $\gamma(1+\beta \cos\theta)\omega_{p(S)}$, bandwidth  $\gamma(1+\beta \cos\theta)\Gamma_{p(S)}$, pulse duration  $T_{p(S)}/(\gamma(1+\beta \cos\theta))$, and laser peak intensity  $\gamma^{2}(1+\beta \cos\theta)^{2}I_{p(S)}$. Furthermore, the effective laser intensity has been taken into account $I^{\mathrm{eff}}_{p(S)}=I_{p(S)} \Gamma/(\gamma(1+\beta)\Gamma_{p(S)})$, with $\Gamma$ the nuclear transition width and $\Gamma_{p(S)}$ the laser bandwidth. 
Further notations used in Eq.~(\ref{eq4}) are  $\epsilon_{0}$ the vacuum permittivity, $\hbar$ the reduced Planck constant, and $\tau_{p(S)}$ the temporal peak position of the pump (Stokes) laser, respectively.

In the following we address the laser beam parameter requirements. 
The most important prerequisite for nuclear STIRAP is the $temporal$ coherence of the x-ray lasers.  The coherence time of the existent XFEL at the Linac Coherent Light Source (LCLS) in Stanford, USA and of the European XFEL are on the order of 0.2~fs, much shorter than the pulse duration of 100~fs  \cite{slac,slac.nature,xfel}. The SXFEL, considered as an upgrade for both facilities, will deliver completely transversely and temporally coherent pulses, that can reach 0.1~ps pulse duration and about 10 meV bandwidth \cite{sxfel.Saldin,LCLSII}. Another option is the XFELO that will provide  coherence time on the order of the pulse duration $\sim$~1~ps, and meV narrow bandwidth  \cite{Kim2008}.  We consider here the laser photon energy for the pump laser fixed at 25 keV for the XFELO and 12.4 keV for the SXFEL. 
The relativistic factor $\gamma$ is given by the resonance condition $E_{3}-E_{1}=\gamma(1+\beta)\hbar\omega_{p}$. The frequency of the Stokes x-ray laser  can be then determined depending on the geometry of the setup. For copropagating  pump and Stokes beams (implying a two-color XFEL), the photon energy of the Stokes laser is smaller than that of the pump laser since  $E_2>E_1$. The alternative that we put forward is to consider two crossed laser beams generated by a single-color SXFEL meeting the accelerated nuclei as shown schematically in Fig.~\ref{fig1}(b). The angle  $\theta_{s}$ between the two beams is determined such that in the nuclear rest frame the pump and Stokes photons fulfill the resonances with two different nuclear transitions. The values of 
$\gamma$, $E_S$ and $\theta_S$ for NCPT for the nuclear systems under consideration are given in Table~\ref{table1}.
The separation of the pump and Stokes beams out of the original XFEL beam requires dedicated x-ray optics such as the diamond mirrors  \cite{YuriV.Shvyd'ko2010,Lindberg2011} developed for the XFELO. X-ray reflections can also help tune the intensity of the two beams. 

The relative coherence between the two ground states is crucial for successful NCPT via STIRAP. Since in our case the lifetime of $\vert2\rangle$ is much longer than the laser pulse durations, decoherence is related to the unstable central frequencies  of the pump and Stokes lasers.  Our single-color XFEL crossed-beam setup accommodates the present lack of two-color x-ray coherent sources
 (only expected as a further upgrade of the LCLS \cite{LCLSII}) 
and reduces the effect of laser central frequency jumps to equal detunings in the pump and Stokes pulses. 
Variations in detuning up to  $\Delta_p=\Delta_S=$10~meV lead to less than 5$\%$ decrease in NCPT. One should mention however that due to time dilation and pulse delay,  a phase jump in the original x-ray beam does not act simultaneously on the pump and Stokes laser in the nuclear rest frame. Coherent population transfer in our setup therefore still requires temporal coherence for the whole pulse duration, as predicted for both SXFEL and XFELO.

\begin{figure}[!]

\center{
  \includegraphics[width=8cm]{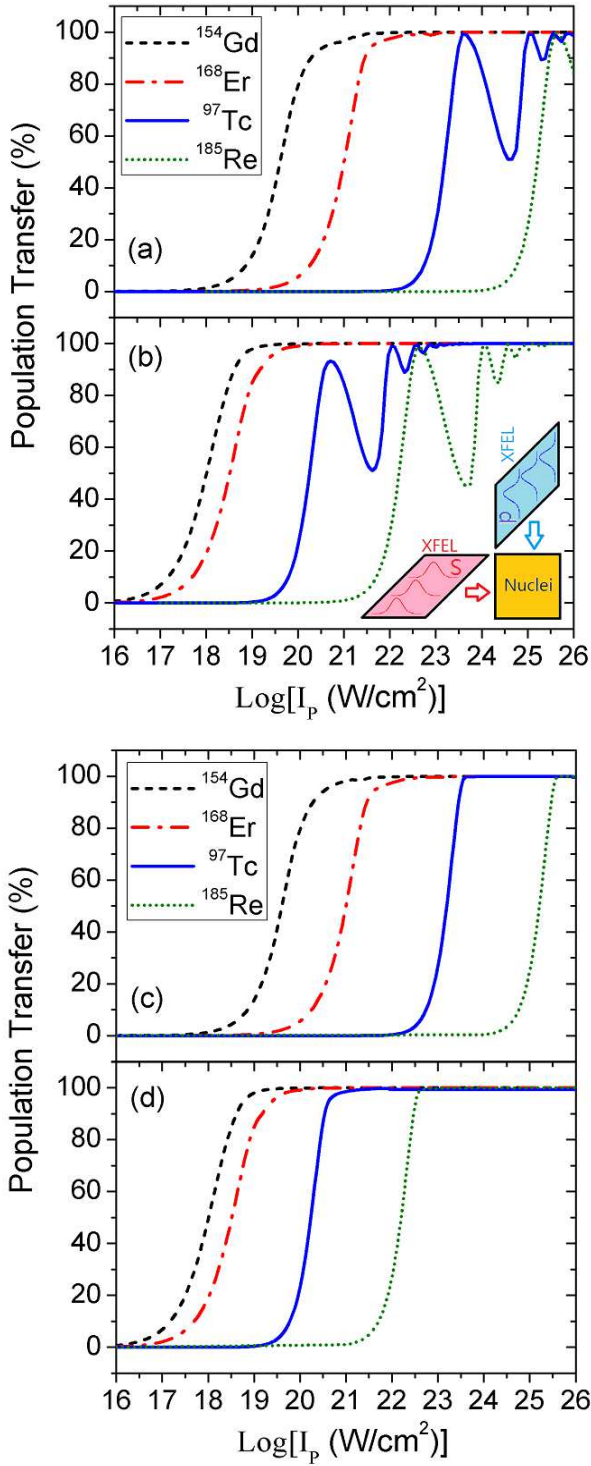}}

 \caption{NCPT for several nuclei as a function of the  pump laser intensity using  SXFEL (a,c) and  XFELO (b,d) parameters. For the crossed-beams setup (a) and (b), the Stokes laser intensities were chosen $I_{S}=0.02 I_{p}$ for $^{185}\mathrm{Re}$,
$I_{S}=0.34 I_{p}$ for $^{168}\mathrm{Er}$, $I_{S}=0.81 I_{p}$ for $^{154}\mathrm{Gd}$ and $I_{S}=20.82 I_{p}$ for $^{97}\mathrm{Tc}$, respectively, according to the $\pi$ pulse intensity ratios $I^{\pi}_{S}/I^{\pi}_{p}$. The inset in (b) depicts the wave-front form necessary to extend the spatial overlap region of the laser and ion beams where STIRAP may occur.
 In the two-color setup (c) and (d), $I_{S}=0.03 I_{p}$ for $^{185}\mathrm{Re}$, $I_{S}=0.35 I_{p}$ for $^{168}\mathrm{Er}$, $I_{S}=0.90 I_{p}$ for $^{154}\mathrm{Gd}$ and $I_{S}=35.06 I_{p}$ for $^{97}\mathrm{Tc}$. All detunings are $\bigtriangleup_{p}=\bigtriangleup_{S}=0$. See discussion in the text and Tables~\ref{table1} and \ref{table2} for further parameters.
  }\label{fig2}
\end{figure}

In Fig.~\ref{fig2} we compare our calculated population transfer for several  cases in both regimes $(i)$ and $(ii)$ using  SXFEL (Fig.~\ref{fig2}(a)) and XFELO (Fig.~\ref{fig2}(b)) parameters in a crossed-beam single-color XFEL setup. For the two-color copropagating beams setup, the results  using SXFEL and  XFELO parameters are showed in Fig.~\ref{fig2}(c) and Fig.~\ref{fig2}(d), respectively. We investigate first the efficiency of NCPT for nuclear three-level systems that do not present a metastable state. The considered nuclear transition energies, multipolarities and reduced matrix elements are given in Table~\ref{table2}.
The choice of nuclei is related to  nuclear data availability and the lifetime values of state $|3\rangle$ required by the two parameter regimes $(i)$ and $(ii)$.  The optimal set of laser parameters is obtained by a careful analysis of the dependence between pump peak intensity $I_{p}$ and pulse delay $\Delta\tau=\tau_{p}-\tau_{s}$. 
A negative time delay corresponds to the $\pi$-pulse population transfer regime, while a positive one stands for STIRAP. 
For each value of $I_{p}$, the $\tau_{p}-\tau_{s}$ is chosen such that the NCPT reaches its maximum value. 

\begin{table}[h]
\vspace{-0.4cm}
\caption{\label{table2}
Nuclear parameters. $E_{i}$ is the energy of state $\vert i \rangle$ with $i\in \{1,2,3\}$ (in keV) \cite{NuclearDataBase}, The initial state $|1\rangle$ is the ground state except for $^{97}$Tc where originally the isomeric state at $E_1=96.57$~keV is populated. The multipolarities and reduced matrix elements (in Weisskopf units, wsu) for the transitions $|3\rangle \rightarrow |j\rangle$ with  $j\in \{1,2\}$ specified in the column header are also given.
}
\center{
\begin{tabular}{lllcccc}
\hline
& & &   \multicolumn{2}{c}{$\epsilon/\mu L$} &
\multicolumn{2}{c}{$B(\varepsilon/\mu\, L)$ (wsu)} \\
\cline{4-7}
Nucl.&  $E_{3}$& $E_{2}$&  $| 1\rangle$
&  $ |2\rangle$ &  $| 1\rangle$  & $ |2\rangle$ \\
\hline
$^{185}$Re &284&125&E2&M1&$6.4\times 10$&$3.7\times 10^{-1}$\\
$^{97}$Tc &657&324&E2&E1&$5\times 10^{2}$&$6.7\times 10^{-5}$\\
$^{154}$Gd &1241&123&E1&E1&$4.4\times 10^{-2}$&$4.9\times 10^{-2}$\\
$^{168}$Er &1786&79&E1&E1&$3.2\times 10^{-3}$&$9.1\times 10^{-3}$\\
\hline
\end{tabular}
}

\end{table}

For regime $(i)$ that allows NCPT via both $\pi$ pulses and STIRAP, we considered the lowest three nuclear levels of $^{185}$Re.  
In the crossed-beam setup, NCPT is achieved at lower intensities via sequential $\pi$ pulses. At the exact $\pi$-pulse value of the pump intensity, a peak in the nuclear population transfer for $^{185}$Re can be observed, at $I_p=6\times 10^{25}$ W/cm$^{2}$ in  Fig.~\ref{fig2}(a) and $I_p=6\times 10^{22}$ W/cm$^{2}$ in  Fig.~\ref{fig2}(b).   
With increasing $I_p$ in the crossed-beam setup (Fig.~\ref{fig2}(a,b)), the $^{185}$Re nuclei are only partially excited to state $|2\rangle$ and the NCPT yield starts to oscillate. The amplitude and frequency of the oscillations are varying as a result of our pulse delay optimization procedure.  At sufficient  intensities in the pulse overlap regime STIRAP becomes preferable as compared to the $\pi$ pulses mechanism due to the lack of oscillations. The plateau at 100$\%$ population transfer indicates that NCPT via STIRAP alone is reached. 
In the two-color copropagating beams scheme (Fig.~\ref{fig2}(c,d)), the pulse shape of pump and that of Stokes are the same in the nuclear rest frame. This renders STIRAP more efficient and thus preferable  compared to the single-color setup, as the STIRAP plateau can be reached  with lower laser intensities.

For  case $(ii)$, we present our results for  $^{154}$Gd and $^{168}$Er,  that require stronger nuclear acceleration with $\gamma$ factors between 24 and 72 and fs pulse delays.  
The $^{154}$Gd ground state population starts to be coherently channeled at  about $I_{p}=10^{17}$ W/cm$^{2}$ using XFELO and $I_{p}=10^{19}$ W/cm$^{2}$ using  SXFEL parameters, respectively. Up to $I_{p}=10^{19}$ W/cm$^{2}$ (XFELO) and $I_{p}=10^{21}$ W/cm$^{2}$ (SXFEL), more than 95$\%$ of the nuclei reach $\vert 2 \rangle$. 
In this case $\pi$ pulses cannot provide the desired NCPT due to the fast spontaneous decay of state $|3\rangle$ in neither copropagating- nor cross-beam setups. The calculated intensities necessary for complete NCPT are within the designed intensities of the XFEL sources. Considering the operating and designed peak power of 20-100 GW \cite{xfel,slac,sxfel.Saldin,LCLSII} for SXFEL (and about three orders of magnitude less for XFELO) and the admirable focus achieved for x-rays of 7~nm \cite{x-focusing}, intensities could reach as high as 
 $10^{17}-10^{18}$ W/cm$^{2}$ for XFELO \cite{Kim2008} and $10^{21}-10^{22}$ W/cm$^{2}$ for SXFEL \cite{sxfel.Saldin}.
 
One of the most relevant applications of NCPT is isomer pumping or depletion. In Fig.~\ref{fig2} we present our results  for NCPT in $^{97}$Tc nuclei starting from the $E_1=96.57$~keV isomeric state which has a half life of $\tau_1=91$~d.     Like $^{185}$Re, $^{97}$Tc belongs to regime $(i)$ such that NCPT at lower intensities can be achieved via $\pi$ pulses in the crossed-beam setup. The intensity for which complete isomer depletion is achieved  using SXFEL is $I_{p}=4\times10^{23}$ W/cm$^{2}$. Due to the longer pulse duration of the XFELO and consequently higher losses via spontaneous decay of state $|3\rangle$, the peak population transfer at $I_p=5.2\times 10^{20}$ W/cm$^{2}$ reaches only 93$\%$ in Fig.~\ref{fig2}(b) in the crossed-beam setup. For the copropagating beams setup, 100$\%$ NCPT is achieved for the same intensity  $I_p=5.2\times 10^{20}$ W/cm$^{2}$.
Compared to the case of high-energy nuclear transitions $(ii)$, the intensities required for isomer depletion are in this case larger, mainly due to the narrow transition width of state $|3\rangle$. Typically, triggering levels high above isomeric states, that would present the advantage of larger linewidths, are less well known. A detailed  analysis of nuclear data in the search for the best candidate is therefore required for successful isomer depletion.

NCPT is sensitive to the fulfillment of the resonance condition. This involves on the one hand precise knowledge of the nuclear transition energy and on the other hand good control of  laser frequency and therefore nuclear acceleration. The former is usually attained in nuclear forward scattering by scanning first for the position of the nuclear resonance. In our setup,  the relativistic factor $\gamma$ influences the detunings and the effective pump and Stokes intensities and Rabi frequencies. For narrow-width excitations $(i)$ it is necessary to first find the laser bandwidth window of the nuclear transition, since most of the transition energy values are not known with such precision. Once found, our procedure of considering an effective intensity which is scaled according to the number of resonant photons should provide the correct approach for a zero-detuning situation.  For the case $(ii)$ where the MeV nuclear transitions have eV widths, it is only necessary to tune the laser photons in the corresponding energy window.

Ion accelerators to bridge the gap between nuclear transition and x-ray laser energies are an important ingredient for achieving NCPT. In the low $\gamma$ region, the forthcoming FAIR at GSI will provide high quality ion beams with energies up to 45 GeV/u \cite{FAIRGSI2006}. The corresponding $\gamma$ limit is about 48 and the precision $\Delta E/E\sim 2\times10^{-4}$. For the high $\gamma$ region, the Large Hadron Collider (LHC) is currently the only suitable ion accelerator which can accelerate $^{208}$Pb$^{82+}$ up to $\gamma=2963.5$ with low energy spread of about $10^{-4}$ \cite{LHC21}. LHC  can also accelerate lighter ions to energies larger than 100 GeV \cite{Carminati2004}. 
For the strong acceleration regime, the resonance condition corresponds to an energy spread of the ion beam of 
$10^{-5}$. This issue becomes more problematic for NCPT of nuclei in the moderate acceleration regime where the resonance condition requires a more precise $\gamma$ value, $\Delta \gamma/\gamma=10^{-6}$. 
On the other hand, the European XFEL will deliver laser pulses with a divergence angle of about $10^{-6}$ rad \cite{xfel}.  This causes the mismatch of $\Delta_p\neq\Delta_S$ together with the energy spread $\Delta E$ of the ion beam. We find NCPT maintains a value of around $80\%$ in the region of $\Delta\theta_{s}=\pm 10^{-5}$ rad and $\Delta \gamma/\gamma=\pm 10^{-6}$ for $^{154}$Gd and $^{168}$Er. This can be compensated by increasing the laser intensity by a factor of three to obtain $100\%$ NCPT.

A further study of the overlap efficiency for the laser beams and ion bunches shows that the copropagating laser beams setup is more advantageous. Using LHC beam size parameters \cite{LHC21} and a 10~$\mu$m radius of the XFEL focusing spot, we estimate that for copropagating laser beams up to $3\times10^5$ nuclei meet the laser focus per bunch and laser pulse, while for crossed laser beams this number reduces to 80 for the smallest overlap volume at $\theta_S=90^{\circ}$. The extreme temporal and spatial fine-tuning required to match the overlaps of a bunched ion beam with the two laser beams in the crossed-beam setup is however at present  challenging. A continuous ion beam, on the other hand, has the disadvantage of much lower ion density at the overlap with the pump and Stokes beam and of no possibility to control when  the ions pass through the overlap region. Furthermore, the necessary time delay between pump and Stokes and the adiabaticity condition $\Omega_{\rm{eff}}\Delta \tau\gg 1$ \cite{K.Bergmann1998} for STIRAP will be in this case only fulfilled for ions at the diagonal line of the overlap area. In order to maintain the pulse delay and the adiabaticity condition for the whole overlap region with the nuclear beam, a special laser pulse front as presented in the inset  Fig.~\ref{fig2}(b) is required. With optical lasers, such a design can be achieved with the help of dispersive glass or specially-shaped mirrors, that could also be developed for x-rays \cite{Shvydko_book}.  We conclude therefore that for a number of technical and conceptual difficulties, the two-color copropagating beams scheme might have better chances to be realized experimentally in the near future.

X-ray coherent light sources are not available today at the few large ion acceleration facilities. At present  a new materials research center MaRIE (Matter-Radiation Interactions in Extreme) providing both a fully coherent XFEL with photon energy up to 100 keV  and accelerated charged-particle beams is envisaged in the USA \cite{MaRIE}. In addition, the photonuclear physics pillar of the Extreme Light Infrastructure (ELI) in construction in Romania  can provide simultaneously 
a compact XFEL as well as ion acceleration  reaching up to 4-5 GeV \cite{eli}. 
At ELI, the combination of coherent gamma-rays and acceleration of the nuclear target are already under consideration 
for  nuclear resonance fluorescence experiments \cite{eli}. Furthermore, ELI is also envisaged to deliver coherent gamma rays with energies of few MeV \cite{eli}, which could be used for direct photoexcitation of giant dipole resonances 
\cite{Weidenmueller}.

Table-top solutions for both ion acceleration and x-ray coherent light would facilitate the experimental realization of isomer depletion in NCPT and nuclear batteries. Table-top x-ray  undulator sources are already operational \cite{MatthiasFuchs2009}, with a number of ideas envisaging compact x-ray  FELs \cite{Nakajima2008,Gruener2007}. Rapid progress   spanning five orders of magnitude increase in the achieved light brightness within only two years has been reported \cite{Kneip1,Kneip2}.
In conjunction with the crystal cavities  designed for the XFELO, such table-top devices have the potential to become a key tool for the release on demand of energy stored in nuclei at large ion accelerator facilities. Alternatively, the exciting forecast of compact shaped-foil-target ion
accelerators \cite{Chen2009} and radiation pressure  acceleration \cite{Bulanov2010} together with microlens beam focusing \cite{microlens} are likely to provide a viable table-top solution to be used together with the existing large-scale XFELs.

In conclusion, the parameter regime for which fully coherent x-ray laser pulses can induce population transfer between nuclear levels  matches the predicted values for the envisaged XFELO and SXFEL facilities. The challenge for the experimental
realization of NCPT and the future of nuclear batteries thus rely especially on the development  of x-ray coherent sources and their conjuncture with ion accelerators,  perhaps making use of high-precision table-top solutions for lasers and ion accelerators to be flexibly used at any location around the globe.

We would like to thank  J\"org Evers, Chang-Yi Wang,  Thorsten Peters, Yen-Wei Lin and Yi-Hsin Chen for fruitful discussions.



\bibliographystyle{elsarticle-num-names}
\bibliography{nstirap}







\end{document}